\documentclass{aa}  

\usepackage{natbib}
\usepackage{graphicx}
\usepackage{txfonts}
\usepackage{hyperref}

\usepackage[switch]{lineno}
\defcitealias{2017MNRAS.465.1621P}{P17}

\begin{document} 
  \title{7.1\,keV sterile neutrino dark matter constraints from a deep \emph{Chandra} X-ray observation of the Galactic bulge Limiting Window.}
  \titlerunning{Limiting Window sterile neutrino constraints}
  \author{F. Hofmann\inst{1}
          \and
          C. Wegg\inst{2,1}
          }

  \institute{Max-Planck-Institut f\"ur extraterrestrische Physik, Giessenbachstra{\ss}e, 85748 Garching, Germany\\
              \email{fhofmann@mpe.mpg.de}
         \and
             Université Côte d'Azur, Observatoire de la Côte d'Azur, CNRS, Laboratoire Lagrange, France
             }

  \date{Date: ...}

  \abstract
  {Recently an unidentified emission line at 3.55\,keV has been detected in X-ray spectra of clusters of galaxies. The line has been discussed as a possible decay signature of 7.1\,keV sterile neutrinos, which have been proposed as a dark matter (DM) candidate.}
   {We aim to further constrain the line strength and its implied mixing angle under the assumption that all DM is made of sterile neutrinos.}
   {The X-ray observations of the Limiting Window (LW) towards the Galactic bulge (GB) offer a unique dataset for exploring DM lines. We characterize the systematic uncertainties of the observation and the fitted models with simulated X-ray spectra. In addition we discuss uncertainties of indirect DM column density constraints towards the GB to understand systematic uncertainties in the assumed DM mass in the field of view of the observation.}
   {We found tight constraints on the allowed flux for an additional line at 3.55\,keV with a positive ($\mathrm{\sim1.5\sigma}$) best fit value $\mathrm{F_X^{3.55keV}\approx(4.5\pm3.5)\times10^{-7}~cts~cm^{-2}~s^{-1}}$. This would translate into a mixing angle of $\mathrm{sin^{2}(2\Theta)\approx(2.3\pm1.8)\times10^{-11}}$ which, while consistent with some recent results, is in tension with earlier detections.}
   {We used a very deep dataset with well understood systematics to derive tight constraints on the mixing angle of a 7.1\,keV sterile neutrino DM. The results highlight that the inner Milky Way will be a good target for DM searches with upcoming missions like eROSITA, XRISM, and ATHENA.}


\maketitle


\section{Introduction}

The Limiting Window (LW) is located 1.5\,deg south of Sgr\,A* (in Galactic coordinates). With its comparatively low foreground absorption ($\mathrm{n_H\lesssim10^{22}~cm^{-2}}$) it was used to resolve about 80 per cent of the Galactic centre hard X-ray emission into point sources and study the source population \citep[][]{2009Natur.458.1142R}. 

\citet{2014ApJ...789...13B} and \citet{2014PhRvL.113y1301B} recently found indications for a weak unidentified emission line ($\rm{E \sim 3.55~keV}$) in X-ray CCD spectra of the Andromeda galaxy and in deep observations of clusters of galaxies using \emph{Chandra} and \emph{XMM-Newton} data.
The line has been proposed as a candidate dark matter (DM) decay line and could be explained by the decay of sterile neutrinos with a mass of $\rm{m_s\approx7.1~keV}$. In the model they can decay into an X-ray photon with $\rm{E_{\gamma}=m_s/2}$ and an active neutrino $\rm{\nu}$. Sterile neutrinos with masses in the keV range have long been discussed as a possible component of DM \citep[e.g.][]{1994PhRvL..72...17D, 2001ApJ...562..593A, 2009ARNPS..59..191B}, but up until recently only upper limits could be derived \citep[e.g. from observations of the Andromeda galaxy or the Bullet Cluster by][]{2008MNRAS.387.1361B, 2008ApJ...673..752B}.
A detection of the sterile neutrino decay would help to discriminate between different production mechanisms \citep[see e.g.][]{2015PhLB..749..283M,2016JKPS...69.1375K,2018JCAP...06..036H}. For a review on the current state of constraints see \citet{2019PrPNP.104....1B}.

In the case of sterile neutrino decay the measured additional flux at $\rm{\sim3.55~keV}$ would be related to two defining properties of the particles: The particle mass $\rm{m_s}$ and the mixing-angle $\rm{sin^2(2\Theta)}$, which describes interaction of the sterile neutrinos with their active neutrino counter-parts and thus the likelihood of decay in the $\rm{\gamma/\nu}$ channel. They are related through \citep[adapted from][]{2014ApJ...789...13B}

\begin{equation}    
 \frac{\sin^2(2\Theta)}{10^{-11}} = 3.25 \frac{F_{\rm DM}}{0.1 {\rm cm}^{-2} {\rm s}^{-1} {\rm sr}^{-1}}  \frac{10^9 {\rm M_\odot} {\rm kpc}^{-2} }{S_{\rm DM}}  \left( \frac{7\,{\rm keV}}{m_s} \right)^4 
\label{eqn:sin}
\end{equation}

where $F_{\rm DM}$ is the observed flux of the DM decay line and $S_{\rm DM}=\int \rho_{\mathrm {DM}}\,dr$ is the DM column density.

We used deep archival X-ray observations of the LW with the \emph{Chandra} telescope to constrain the allowed additional flux in the 3.55\,keV range and the strength of the proposed decay line in the direction of the Galactic bulge (GB). 

The paper is structured as follows: 
Sect.\,2 describes the data used and how it was analysed from reduction to spectral fitting. Sect.\,3 describes how the DM mass in the FOV was derived. Sect.\,4 describes the obtained constraints on flux and mixing angle in the sterile neutrino DM case. In Sect.\,5 we discuss the results in the context of previous work, and Sect.\,6 summarizes the conclusions of the analysis and gives an outlook for future missions like eROSITA.
Uncertainties are quoted on the $\mathrm{1\sigma}$ level unless stated otherwise. 
Abundances are according to solar abundances as in \citet{1989GeCoA..53..197A}.

\section{X-ray data reduction and analysis}

We used all available archival observations of the LW with the \emph{Chandra} Advanced CCD Imaging Spectrometer \citep[ACIS;][]{2003SPIE.4851...28G} using the imaging (ACIS-I) CCD array (about 0.1 to 10\,keV energy range). This instrument provides high spatial ($\rm{\sim1\arcsec}$) and spectral resolution ($\rm{\sim100~eV}$ full width half maximum, FWHM).
We used the observations with ObsIDs: 5934, 6362, 6365, 9500, 9501, 9502, 9503, 9504, 9505, 9854, 9855, 9892, and 9893.
The observations were reprocessed using the \emph{Chandra} Interactive Analysis of Observations software package \citep[CIAO;][]{2006SPIE.6270E..1VF} version 4.5 and the \emph{Chandra} Calibration Database \citep[CalDB;][]{2007ChNew..14...33G} version 4.5.9.

The data analysis is based on \citet{2016A&A...585A.130H,2016A&A...592A.112H} but we describe the most important steps again in the following.
We merged all ACIS-I observations of the LW, removed detected point sources, and extracted a spectrum from the remaining $\mathrm{66~arcmin^2}$ \citep[see also technique by][]{2006MNRAS.371..829S}.
The background was extracted from a matched "blank-sky" observation (using \texttt{acis\_bkgrnd\_lookup}) and renormalised to match the count rate of the source spectrum in the 10.0-12.5\,keV energy range.
The \emph{Chandra} "blank-sky" background is created from observations at least 20\,deg from the Galactic plane. This could potentially bias our results low by about ten per cent (see profile Fig. \ref{fig:prof}), because we might be subtracting some of the line strength with the background.
The response files were averaged and weighted by the number of counts in the spectrum (both auxiliary response files, ARF, and redistribution matrix files, RMF). For analysing the spectra we used \texttt{XSPEC} version 12.9.1u \citep[][]{1996ASPC...99..409A} and ATOMDB version 3.0.7 \citep[][]{2012ApJ...756..128F}.

To estimate the upper limit of the flux allowed for an additional emission line, we searched for the best fitting \texttt{apec} model (with two temperature components) for collisionally-ionized plasma with absorption ($\rm{n_H}$) and an additional zero-width Gaussian line.
The normalisation of the Gaussian was allowed to be negative to avoid bias.
The spectrum was grouped to contain a minimum of 22 raw counts in each bin (using \texttt{grppha}) and we used the range from 2-5\,keV for fitting the spectral model to the data (using $\rm{\chi^2}$ statistics). Free parameters of the fit were the normalisation of the spectral components, the temperatures, and the relative abundances of the \texttt{apec} models. 

Once the best fit was identified, we calculated the confidence intervals (99.7 per cent) for the additional flux added by the Gaussian using a Monte Carlo Markov Chain (MCMC) with length of $\mathrm{10^{4}}$ and burn-in length of $\mathrm{10^{3}}$.
The average best fit parameters of the \texttt{apec} models are (left free for each individual fit): foreground absorption by neutral hydrogen $\mathrm{n_H \approx 0.5 \times 10^{22}~cm^{-2}}$ (same for both components), temperature of the first model: $\mathrm{kT_1 \approx 0.9~keV}$, second temperature: $\mathrm{kT_2 \approx 0.7~keV}$, metallicity: $\mathrm{Z \approx 0.7~Z_\odot}$ (same for both models), and goodness of fit: $\mathrm{\chi^2_{red.} \approx 1.2}$ \citep[consistent with \emph{Suzaku} measurements in the region by][]{2013ApJ...773...20N}.

\section{Dark matter mass model}

Driven by data from the Gaia satellite, over the coming years the distribution of dark matter within the Milky Way will be clarified. Presently however,  the dark matter profile of the Milky Way remains uncertain, particularly in the Baryon dominated inner regions which the LW probes \citep[for a review see][]{2016ARA&A..54..529B}. 
The most direct estimates of the stellar surface density in the direction of the LW come from measurements of the microlensing optical depth towards the Bulge \citep{2016MNRAS.463..557W}. Our dark matter profile leaves sufficient baryonic mass remaining from the mass budget allowed by the Galactic rotation curve to satisfy these microlensing constraints.

Our fiducial dark matter profile is that found in \citet[][hereafter \citetalias{2017MNRAS.465.1621P}]{2017MNRAS.465.1621P} by fitting triaxial dynamical models of the barred bulge to a range of photometric and spectroscopic data on the stars in the inner Galaxy. These models constrained the amount of dark matter in the central 2kpc of the Galaxy, and \citetalias{2017MNRAS.465.1621P} found that to also simultaneously match the Galactic rotation curve and local constraints, a cored Einasto profile was preferred. The best fitting model from \citetalias{2017MNRAS.465.1621P} has a dark matter profile which is flattened with axis ratio $q=0.8$ and can be parameterized using the ellipsoidal radius $m=\sqrt{x^2 + y^2 + (z/q)^2}$ as
\begin{equation}
\rho_{\rm dm} = \rho_{\rm 0} \exp \left\{ -\frac{2}{\alpha}\left[ \left(\frac{m}{m_0}\right)^\alpha - 1\right]\right\}
\end{equation}
with  $\rho_{\rm 0} = 0.018\,\mathrm{M_{\odot}~pc^{-3}}$, $m_0 = 7.1\,{\rm kpc}$, $\alpha=0.77$. This corresponds to a dark matter density at the Sun of $\rho_{\rm dm,\odot}=0.013\,\mathrm{M_{\odot}~pc^{-3}}=0.50\,\mathrm{GeV~c^{-2}~cm^{-3}}$, consistent with recent estimates \citep[e.g.][]{2014JPhG...41f3101R}. 

In the LW this model corresponds to a column density of $S_\mathrm{DM} = 1.1\times10^9\,\mathrm{M_\odot}\,\mathrm{kpc}^{-2}$. However, because of the uncertainty in the Milky Way's DM profile, we discuss later in Sect.\,4 and 5 the freedom there remains to alter these column densities.

\section{Flux and mixing angle constraints}

\begin{figure}
    \centering
    \includegraphics[width=0.49\textwidth]{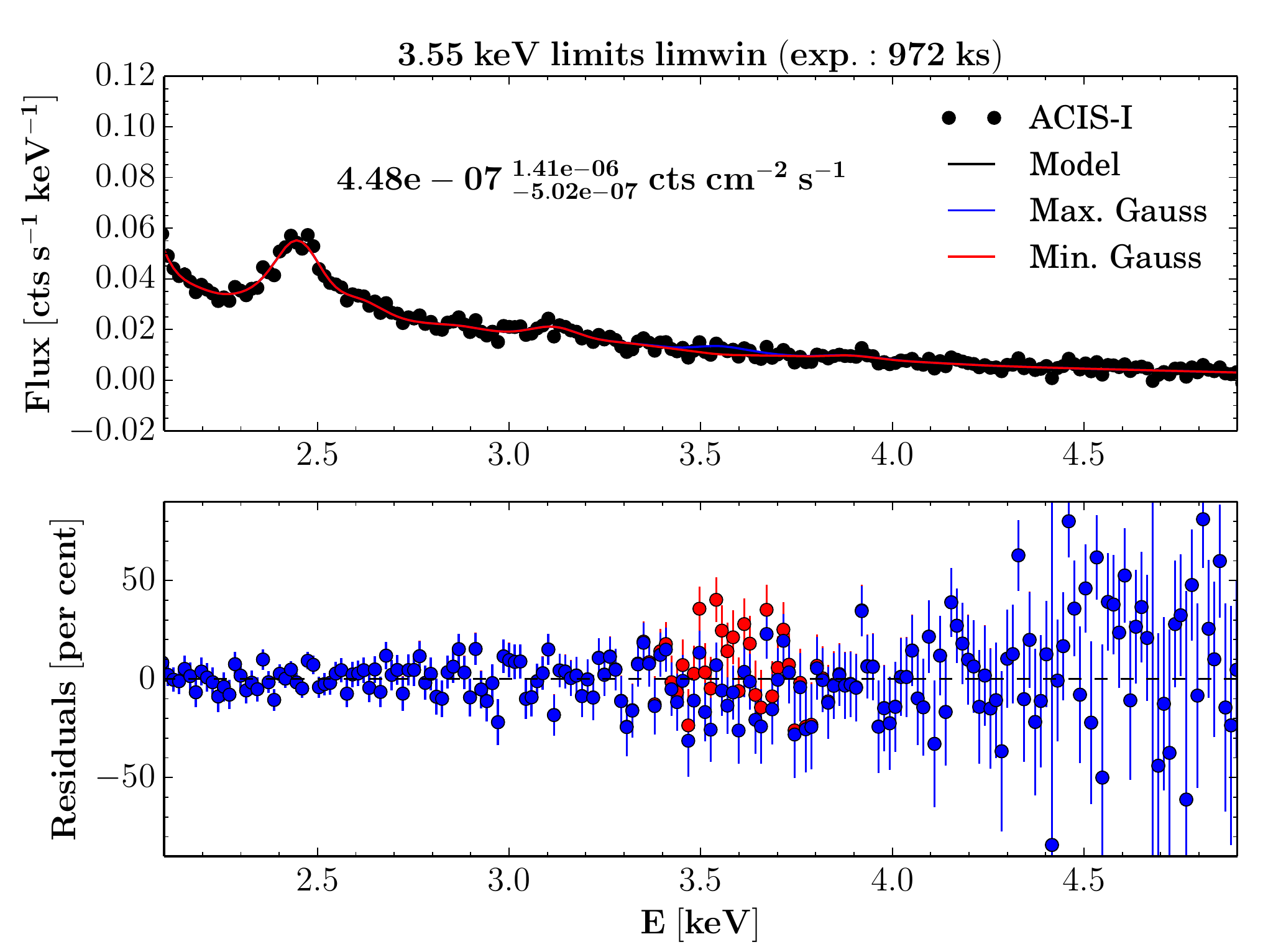}
    \caption{Limiting Window $\mathrm{\sim1~Ms}$ \emph{Chandra} spectrum with constraints on an additional 3.55\,keV Gaussian line on top of a standard two-temperature \texttt{apec} model. Uncertainties are given at 99.7 per cent confidence range ($\mathrm{\sim3\sigma}$). Residuals are given in per cent deviation from the model. Red residuals are for the model with minimum allowed flux and blue for the model with maximum flux in the additional Gaussian line.}
    \label{fig:355}
\end{figure}

We obtained a $\mathrm{3\sigma}$ upper flux limit of $\mathrm{\sim1.5\times10^{-6}~cts~cm^{-2}~s^{-1}}$ (see Fig. \ref{fig:355}) which translates to a sterile neutrino mixing angle upper-limit of $\mathrm{sin^2(2\Theta)\lesssim7.7\times10^{-11}}$, assuming all DM is made up of 7.1\,keV sterile neutrinos.
The best fit values and $\mathrm{1\sigma}$ uncertainties are $\mathrm{F_X^{3.55keV}\approx(4.5\pm3.5)\times10^{-7}~cts~cm^{-2}~s^{-1}}$ and\\ $\mathrm{sin^2(2\Theta)\approx(2.3\pm1.8)\times10^{-11}}$.

\begin{figure}
    \centering
    \includegraphics[width=0.50\textwidth]{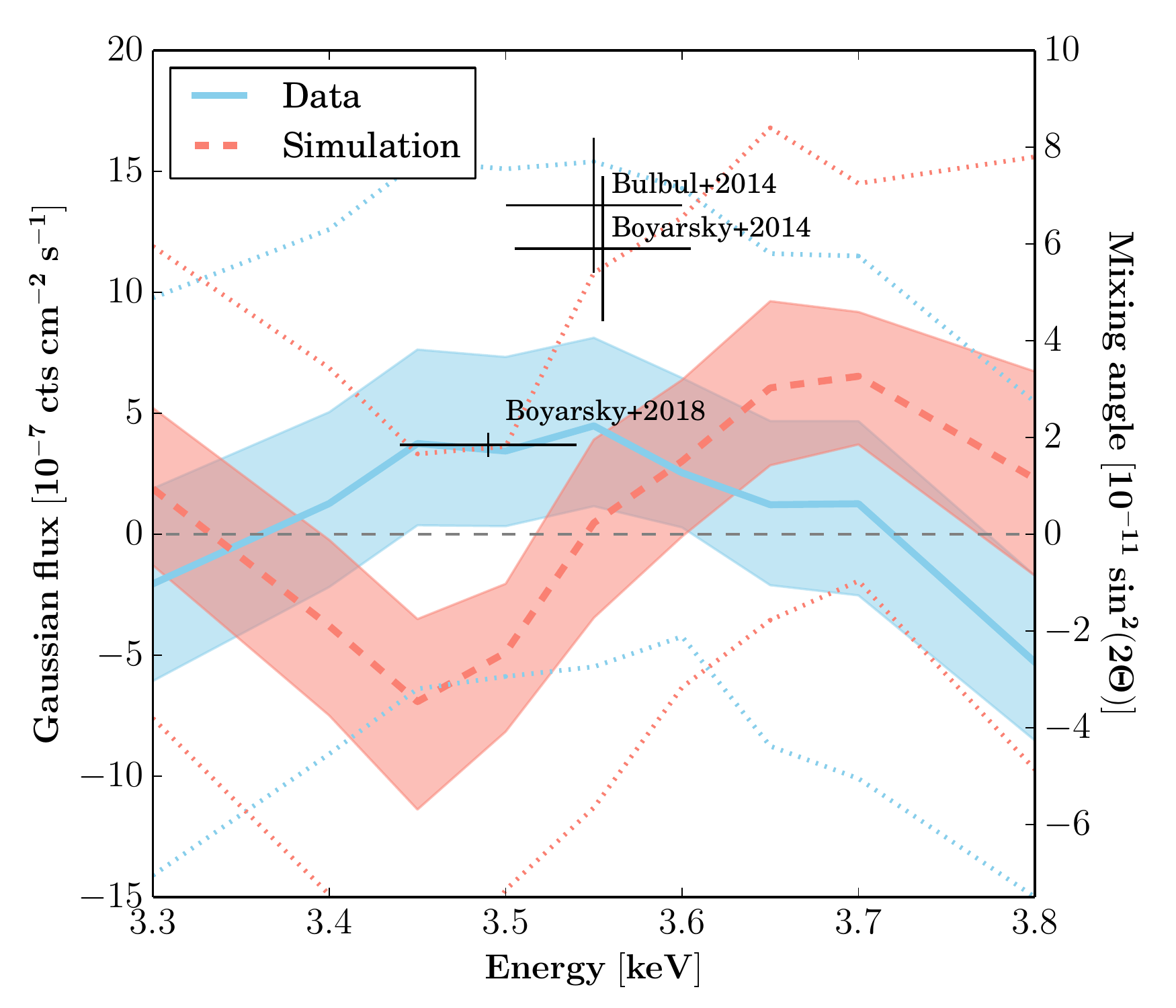}
    \caption{The solid line and shaded region show the best-fit and $\mathrm{1\sigma}$ limits on additional Gaussian flux in the Limiting Window X-ray spectrum at different energies. The dashed line and shaded region show the results for a simulated spectrum without any additional line. The $\mathrm{3\sigma}$ uncertainties are plotted as dotted lines. The black crosses show the expected flux from selected previous detections of the line with $\mathrm{1\sigma}$ uncertainties in flux and the typical energy resolution of the instruments ($\mathrm{\sim100~eV}$). The second y axis shows the corresponding mixing angle.}
    \label{fig:step}
\end{figure}

Fig. \ref{fig:355} shows the best fit and limits for the additional Gaussian line at 3.55\,keV. Fig. \ref{fig:step} and Tab. \ref{tab:fit} show the best-fit and limits in flux for a range of energies between the two stronger line complexes of Sulfur (S) and Argon (Ar) at about 3.0-3.2\,keV and Calcium (Ca) at about 3.8-4.0\,keV.
In addition we show best-fit and limits derived from a simulated spectrum with the same properties, but no line at 3.55\,keV, analysed in the same way.
Fluctuations of $\mathrm{1-2\sigma}$ appear also in the simulation, but have a different form and are overall zero in the analyzed range. In contrast the real data shows a continuous increase towards the energy of 3.55\,keV.
Fig. \ref{fig:step} shows the expected flux, scaling the detections of \citet{2014PhRvL.113y1301B} in M\,31, \citet{2014ApJ...789...13B} in clusters of galaxies, and \citet{2018arXiv181210488B} in Galactic halo observations with \emph{XMM-Newton}. 

Systematic uncertainties in the DM mass models but also underestimated plasma emission lines in the 3.55\,keV range could cause the observed tension.
There is good agreement of the \emph{XMM-Newton} measurement of \citet{2018arXiv181210488B} (see Fig. \ref{fig:step}).
The offset to the values from \citet{2014ApJ...789...13B} and \citet{2014PhRvL.113y1301B} could be explained if the normalization of the Galactic DM profile was a factor of about two lower than the estimate from \citetalias{2017MNRAS.465.1621P}. This lies at the boundary of conceivable column densities in the Milky Way: even using a relatively low dark matter contribution to the circular velocity near the Sun of $V_{c,{\rm dm}} \approx 100\,{\rm km}\,{\rm s}^{-1}$ \citep{2013ApJ...779..115B} and making the extreme assumption that this mass is a constant density sphere only reduces the dark matter column density towards the LW by a factor $\sim 2$ to $\approx 0.6\times10^9\,\mathrm{M_\odot}\,\mathrm{kpc}^{-2}$. This is because our fiducial dark matter profile already has a $\sim$kpc size core, meaning there is relatively little freedom to increase mixing angle by reducing the dark matter column density. 
The tension with \citet{2014PhRvL.113y1301B} in M\,31 may be reduced by different dark matter mass modelling in which considerable uncertainty remains \citep[e.g. using the new models of M\,31 produced by][]{2018MNRAS.481.3210B}.

\begin{table}[h]
\caption[]{Additional Gaussian flux allowed (fit in 2.5-5.0\,keV range) at different line energies.}
\begin{center}
\begin{tabular}{llll}
\hline\hline\noalign{\smallskip}
  \multicolumn{1}{l}{$\mathrm{E_{line}}$[keV]} &
  \multicolumn{1}{l}{Best fit\tablefootmark{a}} &
  \multicolumn{1}{l}{Upper limit} &
  \multicolumn{1}{l}{Lower limit} \\
\noalign{\smallskip}\hline\noalign{\smallskip}
  3.30 & -2.05 (1.88) & 9.77 (11.9) & -14.1 (-7.61) \\
  3.40 & 1.27 (-3.79) & 12.6 (6.87) & -9.08 (-14.9) \\
  3.45 & 3.75 (-6.92) & 15.4 (3.32) & -6.39 (-20.3) \\
  3.50 & 3.44 (-4.89) & 15.1 (3.63) & -5.88 (-14.7) \\
  3.55 & 4.48 (0.47) & 15.4 (10.8) & -5.48 (-11.3) \\
  3.60 & 2.54 (3.02) & 14.3 (13.1) & -4.23 (-6.33) \\
  3.65 & 1.22 (6.05) & 11.6 (16.8) & -8.76 (-3.55) \\
  3.70 & 1.26 (6.53) & 11.5 (14.5) & -10.1 (-1.93) \\
  3.80 & -5.29 (2.29) & 5.46 (15.6) & -15.0 (-9.72) \\
\noalign{\smallskip}\hline
\end{tabular}
\tablefoot{
\tablefoottext{a}{Values from MCMC error analysis in XSPEC at $\mathrm{3\sigma}$ confidence level. Flux units are $\mathrm{10^{-7}~cts~cm^{-2}~s^{-1}}$. In brackets are comparison values from simulated spectra without any additional Gaussian emission line and analyzed in the same way to show systematic uncertainties of the method.}
}
\end{center}
\label{tab:fit}
\end{table}

\section{Discussion}

The discussion about the $\mathrm{\sim7~keV}$ sterile neutrino decay line continues with no definite answer in favour or against yet. There have been several recent studies reaching almost $\mathrm{3\sigma}$ exclusions \citep[e.g.][]{2015MNRAS.452.3905A,2016MNRAS.458.3592J,2017ApJ...837L..15A} and studies with detections of about the same significance \citep[e.g.][]{2016PhRvD..94l3504N,2018ApJ...854..179C,2016ApJ...829..124F,2016ApJ...831...55B}. The recent debate suggests that unaccounted systematics in the analysis from instruments, spectral- \citep[][]{2018PASJ...70...12H}, and DM mass-modelling cause these differences. 
Future high spectral resolution instruments like \emph{XRISM} \citep[][]{2018SPIE10699E..22T} will help to reduce systematics in the spectral modelling and instrument calibration.

We were looking for the best archival datasets to constrain the line emission and identified a $\sim1~Ms$ \emph{Chandra} ACIS-I observation of the Limiting Window towards the Galactic bulge among the best. It maximizes the expected DM decay flux, because the Galactic bulge is the closest, highest DM column density object in the sky \citep[][]{2019MNRAS.tmp..674L}. 

In addition the low foreground absorption of $\mathrm{\sim5\times10^{21}~cm^{-2}}$ allows to resolve a large fraction of point sources which could contaminate the spectrum. The remaining $\mathrm{\sim1~keV}$ plasma emission leads to a lower continuum contribution at 3.5\,keV compared to hotter objects like clusters of galaxies. The \emph{Chandra} observations are mostly ($\mathrm{\sim90}$ per cent of time) very long single exposures taken within the year 2008, limiting possible systematics when adding many observations taken over a long period (e.g. better average response approximation).

The presented analysis focused on minimizing systematics by using the latest available spectral models (Sect. 2), a very well understood, deep dataset, and the latest models of the expected DM mass in the FOV.

There have been many attempts to explain the line emission with alternative DM scenarios \citep[e.g.][]{2017PhRvD..96l3009C} or with other unknown processes without the necessity for a DM interpretation \citep[][]{2015A&A...584L..11G,2016ApJ...833...52S}. 
We focused on the DM interpretation, because the HITOMI results \citep[][]{2017ApJ...837L..15A} excluded many other proposed explanations. Charge exchange between different temperature material however remains a viable explanation for at least part of the line strength and would be expected in the LW. The HITOMI high resolution X-ray spectroscopy data of the Perseus cluster was not quite deep enough to constrain the possible DM line if it is broadened by the expected DM velocity $\mathrm{\sim1300~km~s^{-1}}$ \citep[][]{2017ApJ...837L..15A}.

Recent constraints further encouraged soft X-ray observations of the Galactic bulge area as one of the best targets for DM annihilation line searches.
\citet{2019MNRAS.tmp..674L} discussed an overview identifying the highest flux targets for DM decay line searches and \citet{2017PhR...711....1A,2017JCAP...01..025A} summarize the current state of the 3.55\,keV line discussion. At higher energies the existence of unknown lines has been further constrained by \citet{2019PhRvD..99h3005N}.

\begin{figure}
    \centering
    \includegraphics[width=0.49\textwidth]{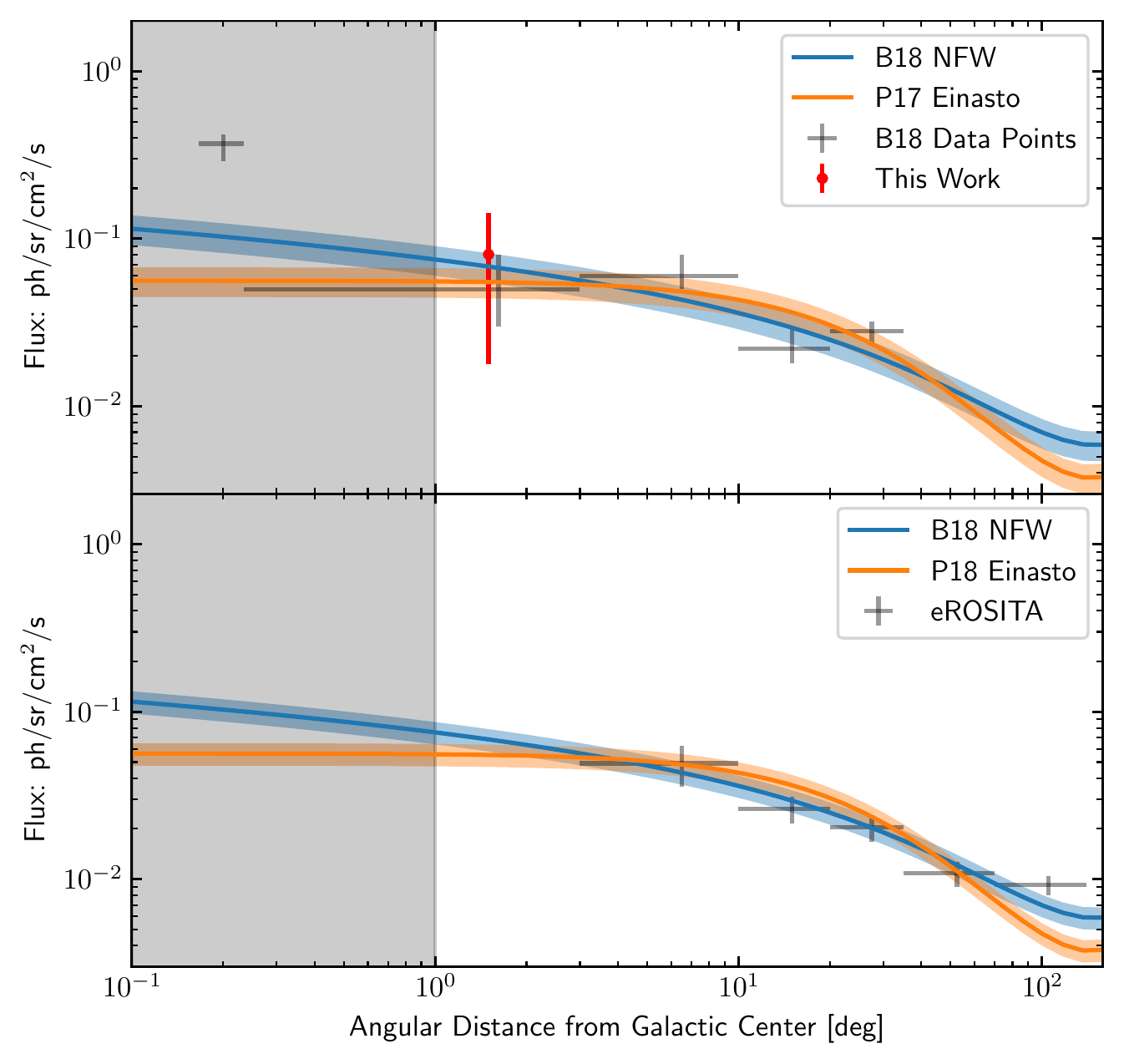}
    \caption{{\bf Upper Panel:} \citet{2018arXiv181210488B}: B18 surface brightness expectations for an NFW DM profile of the Galaxy. \citet{2017MNRAS.465.1621P}: P17 DM profile using B18 decay time of sterile neutrinos. The red data point shows the best fit value with uncertainties from this work. The shaded area indicates where the DM mass models and X-ray measurements become more uncertain. {\bf Lower Panel:} As upper plot, but with data points showing expectations for eROSITA extrapolating from the B18 measurements assuming the errors scale with the root of the relative observation time i.e. that the grasp and background of XMM and eROSITA at 3.5keV are the same. As an all sky survey eROSITA will either accurately measure, or rule out, the 3.5keV line from the wealth of observations far from the Galactic center where the astrophysical backgrounds are low.}
    \label{fig:prof}
\end{figure}

\citet{2018arXiv181210488B} analyzed the surface brightness profile of the line in the halo of the Galaxy which are consistent with our findings within uncertainties. Our 3.55\,keV surface brightness value of $\mathrm{0.09\pm0.05~ph~cm^{-2}~s^{-1}~sr^{-1}}$ agrees very well with their Galactic DM profile and measurements (assuming a quantum efficiency of about 90 per cent for \emph{Chandra} ACIS-I at 3.55\,keV, see Fig. \ref{fig:prof}). 
The constraints in the LW are consistent with the upper limit from \citet{2018arXiv181206976D} in the Galactic halo within uncertainties.

\section{Conclusions and outlook}

   \begin{enumerate}
      \item We used a X-ray dataset in a region with high dark matter column density, but low astrophysical backgrounds due to the high fraction of resolved sources and low gas temperature.
      \item We present a  DM mass model for the analysed field of view which was derived from state-of-the-art dynamical models of the inner Galaxy.
      \item We find some tension with previous measurements of the 3.55\,keV line. The allowed upper limit for the mixing angle in the  7.1\,keV sterile neutrino scenario would be $\mathrm{sin^2(2\Theta)\lesssim7.7\times10^{-11}}$ ($\mathrm{3\sigma}$ confidence level). The $\mathrm{1.5\sigma}$ positive best fit flux would translate to $\mathrm{sin^2(2\Theta)\approx(2.3\pm1.8)\times10^{-11}}$ ($\mathrm{1\sigma}$ uncertainty). 
      \item An alternative explanation of the marginal detections and their tension among each other remain underestimated systematic uncertainties in calibration of the instruments and modelling of the X-ray emission lines.
   \end{enumerate}

The 7.1\,keV sterile neutrino remains one of the best-testable DM candidates for the coming decade especially with new instruments, but there are still possibilities to improve constraints from currently available data.
There is the possibility to expand the study to a recently completed \emph{XMM-Newton} Galactic bulge survey with $\mathrm{\sim500}$ times the FOV area \citep[][]{Ponti2019}, but $\mathrm{\sim40}$ times less average exposure and a more complicated mix of plasma emission as well as much higher contribution from unresolved point sources (more difficult to remove due to lower spatial resolution and shallower exposure). A followup analysis to this work using the additional data might provide improved constraints.

The eROSITA telescope will perform the first all-sky X-ray survey in the 3.5\,keV range \citep[][]{2012arXiv1209.3114M}. 
With its all-sky coverage and comparable grasp at 3.5 keV with respect to \emph{XMM-Newton} we expect to improve the current constraints from Galactic DM halo observations considerably. The largest uncertainty in the ability of eROSITA to constrain the 3.5\,keV line is the level of background. In Fig. \ref{fig:prof} we have  assumed the same background as \emph{XMM-Newton}, but envisage that from it's observation point at L2 (second Lagragian point of the Sun-Earth system) the background might be lower, and therefore performance better than this conservative assumption.

The lower panel of Fig. \ref{fig:prof} shows the expected 3.5\,keV surface brightness profile extrapolating from \citet{2018arXiv181210488B} after the four year survey (scan number eRASS8). 
The XRISM telescope with its high spectral resolution in combination with the eROSITA all-sky coverage will allow to determine the nature of the line and the ATHENA observatory \citep[][]{2013arXiv1306.2307N} will improve constraints even further.

\begin{acknowledgements}

We thank the anonymous referee for comments that helped to improve the clarity of the paper, E. Bulbul and K. Abazajian for helpful discussions, and C. Pulsoni for comments on the manuscript. We acknowledge the use of data obtained from the \emph{Chandra} Data Archive and software provided by the \emph{Chandra} X-ray Center (CXC) in the application package CIAO; NASA's Astrophysics Data System; SAOImage DS9, developed by Smithsonian Astrophysical Observatory; data and/or software provided by the High Energy Astrophysics Science Archive Research Center (HEASARC), which is a service of the Astrophysics Science Division at NASA/GSFC and the High Energy Astrophysics Division of the Smithsonian Astrophysical Observatory; use of the python packages matplotlib, scipy, numpy, and pyXSPEC.
F. Hofmann acknowledges financial support from the BMWi/DLR grant FKZ 50 OR 1715.
C. Wegg  acknowledges  funding  from  the  European Union's Horizon 2020 research and innovation program  under  the  Marie  Sk\l{}odowska-Curie  grant  agreement No 798384.
\end{acknowledgements}

\bibliographystyle{aa} 
\bibliography{references} 

\begin{thebibliography}{49}
\expandafter\ifx\csname natexlab\endcsname\relax\def\natexlab#1{#1}\fi

\bibitem[{{Abazajian} {et~al.}(2001){Abazajian}, {Fuller}, \&
  {Tucker}}]{2001ApJ...562..593A}
{Abazajian}, K., {Fuller}, G.~M., \& {Tucker}, W.~H. 2001, \apj, 562, 593

\bibitem[{{Abazajian}(2017)}]{2017PhR...711....1A}
{Abazajian}, K.~N. 2017, \physrep, 711, 1

\bibitem[{{Adhikari} {et~al.}(2017){Adhikari}, {Agostini}, {Ky}, {Araki},
  {Archidiacono}, {Bahr}, {Baur}, {Behrens}, {Bezrukov}, {Bhupal Dev}, {Borah},
  {Boyarsky}, {de Gouvea}, {Pires}, {de Vega}, {Dias}, {Di Bari}, {Djurcic},
  {Dolde}, {Dorrer}, {Durero}, {Dragoun}, {Drewes}, {Drexlin}, {D{\"u}llmann},
  {Eberhardt}, {Eliseev}, {Enss}, {Evans}, {Faessler}, {Filianin}, {Fischer},
  {Fleischmann}, {Formaggio}, {Franse}, {Fraenkle}, {Frenk}, {Fuller},
  {Gastaldo}, {Garzilli}, {Giunti}, {Gl{\"u}ck}, {Goodman}, {Gonzalez-Garcia},
  {Gorbunov}, {Hamann}, {Hannen}, {Hannestad}, {Hansen}, {Hassel}, {Heeck},
  {Hofmann}, {Houdy}, {Huber}, {Iakubovskyi}, {Ianni}, {Ibarra}, {Jacobsson},
  {Jeltema}, {Jochum}, {Kempf}, {Kieck}, {Korzeczek}, {Kornoukhov},
  {Lachenmaier}, {Laine}, {Langacker}, {Lasserre}, {Lesgourgues}, {Lhuillier},
  {Li}, {Liao}, {Long}, {Maltoni}, {Mangano}, {Mavromatos}, {Menci}, {Merle},
  {Mertens}, {Mirizzi}, {Monreal}, {Nozik}, {Neronov}, {Niro}, {Novikov},
  {Oberauer}, {Otten}, {Palanque-Delabrouille}, {Pallavicini}, {Pantuev},
  {Papastergis}, {Parke}, {Pascoli}, {Pastor}, {Patwardhan}, {Pilaftsis},
  {Radford}, {Ranitzsch}, {Rest}, {Robinson}, {Rodrigues da Silva},
  {Ruchayskiy}, {Sanchez}, {Sasaki}, {Saviano}, {Schneider}, {Schneider},
  {Schwetz}, {Sch{\"o}nert}, {Scholl}, {Shankar}, {Shrock}, {Steinbrink},
  {Strigari}, {Suekane}, {Suerfu}, {Takahashi}, {Van}, {Tkachev}, {Totzauer},
  {Tsai}, {Tully}, {Valerius}, {Valle}, {Venos}, {Viel}, {Vivier}, {Wang},
  {Weinheimer}, {Wendt}, {Winslow}, {Wolf}, {Wurm}, {Xing}, {Zhou}, \&
  {Zuber}}]{2017JCAP...01..025A}
{Adhikari}, R., {Agostini}, M., {Ky}, N.~A., {et~al.} 2017, Journal of
  Cosmology and Astro-Particle Physics, 2017, 025

\bibitem[{{Aharonian} {et~al.}(2017){Aharonian}, {Akamatsu}, {Akimoto},
  {Allen}, {Angelini}, {Arnaud}, {Audard}, {Awaki}, {Axelsson}, {Bamba},
  {Bautz}, {Blandford}, {Bulbul}, {Brenneman}, {Brown}, {Cackett},
  {Chernyakova}, {Chiao}, {Coppi}, {Costantini}, {de Plaa}, {den Herder},
  {Done}, {Dotani}, {Ebisawa}, {Eckart}, {Enoto}, {Ezoe}, {Fabian}, {Ferrigno},
  {Foster}, {Fujimoto}, {Fukazawa}, {Furuzawa}, {Galeazzi}, {Gallo}, {Gandhi},
  {Giustini}, {Goldwurm}, {Gu}, {Guainazzi}, {Haba}, {Hagino}, {Hamaguchi},
  {Harrus}, {Hatsukade}, {Hayashi}, {Hayashi}, {Hayashida}, {Hiraga},
  {Hornschemeier}, {Hoshino}, {Hughes}, {Ichinohe}, {Iizuka}, {Inoue}, {Inoue},
  {Inoue}, {Ishibashi}, {Ishida}, {Ishikawa}, {Ishisaki}, {Itoh}, {Iwai},
  {Iyomoto}, {Kaastra}, {Kallman}, {Kamae}, {Kara}, {Kataoka}, {Katsuda},
  {Katsuta}, {Kawaharada}, {Kawai}, {Kelley}, {Khangulyan}, {Kilbourne},
  {King}, {Kitaguchi}, {Kitamoto}, {Kitayama}, {Kohmura}, {Kokubun}, {Koyama},
  {Koyama}, {Kretschmar}, {Krimm}, {Kubota}, {Kunieda}, {Laurent}, {Lebrun},
  {Lee}, {Leutenegger}, {Limousin}, {Loewenstein}, {Long}, {Lumb}, {Madejski},
  {Maeda}, {Maier}, {Makishima}, {Markevitch}, {Matsumoto}, {Matsushita},
  {McCammon}, {McNamara}, {Mehdipour}, {Miller}, {Miller}, {Mineshige},
  {Mitsuda}, {Mitsuishi}, {Miyazawa}, {Mizuno}, {Mori}, {Mori}, {Moseley},
  {Mukai}, {Murakami}, {Murakami}, {Mushotzky}, {Nakagawa}, {Nakajima},
  {Nakamori}, {Nakano}, {Nakashima}, {Nakazawa}, {Nobukawa}, {Nobukawa},
  {Noda}, {Nomachi}, {O' Dell}, {Odaka}, {Ohashi}, {Ohno}, {Okajima}, {Ota},
  {Ozaki}, {Paerels}, {Paltani}, {Parmar}, {Petre}, {Pinto}, {Pohl}, {Porter},
  {Pottschmidt}, {Ramsey}, {Reynolds}, {Russell}, {Safi-Harb}, {Saito},
  {Sakai}, {Sameshima}, {Sasaki}, {Sato}, {Sato}, {Sato}, {Sawada}, {Schartel},
  {Serlemitsos}, {Seta}, {Shidatsu}, {Simionescu}, {Smith}, {Soong}, {Stawarz},
  {Sugawara}, {Sugita}, {Szymkowiak}, {Tajima}, {Takahashi}, {Takahashi},
  {Takeda}, {Takei}, {Tamagawa}, {Tamura}, {Tamura}, {Tanaka}, {Tanaka},
  {Tanaka}, {Tashiro}, {Tawara}, {Terada}, {Terashima}, {Tombesi}, {Tomida},
  {Tsuboi}, {Tsujimoto}, {Tsunemi}, {Tsuru}, {Uchida}, {Uchiyama}, {Uchiyama},
  {Ueda}, {Ueda}, {Ueno}, {Uno}, {Urry}, {Ursino}, {de Vries}, {Watanabe},
  {Werner}, {Wik}, {Wilkins}, {Williams}, {Yamada}, {Yamaguchi}, {Yamaoka},
  {Yamasaki}, {Yamauchi}, {Yamauchi}, {Yaqoob}, {Yatsu}, {Yonetoku}, {Yoshida},
  {Zhuravleva}, {Zoghbi}, \& {Hitomi Collaboration}}]{2017ApJ...837L..15A}
{Aharonian}, F.~A., {Akamatsu}, H., {Akimoto}, F., {et~al.} 2017, \apj, 837,
  L15

\bibitem[{{Anders} \& {Grevesse}(1989)}]{1989GeCoA..53..197A}
{Anders}, E. \& {Grevesse}, N. 1989, Geochimica et Cosmochimica Acta, 53, 197

\bibitem[{{Anderson} {et~al.}(2015){Anderson}, {Churazov}, \&
  {Bregman}}]{2015MNRAS.452.3905A}
{Anderson}, M.~E., {Churazov}, E., \& {Bregman}, J.~N. 2015, \mnras, 452, 3905

\bibitem[{{Arnaud}(1996)}]{1996ASPC...99..409A}
{Arnaud}, K.~A. 1996, in Cosmic Abundances, ed. S.~S. {Holt} \& G.~{Sonneborn},
  Vol.~99, 409

\bibitem[{{Bla{\~n}a D{\'\i}az} {et~al.}(2018){Bla{\~n}a D{\'\i}az}, {Gerhard},
  {Wegg}, {Portail}, {Opitsch}, {Saglia}, {Fabricius}, {Erwin}, \&
  {Bender}}]{2018MNRAS.481.3210B}
{Bla{\~n}a D{\'\i}az}, M., {Gerhard}, O., {Wegg}, C., {et~al.} 2018, \mnras,
  481, 3210

\bibitem[{{Bland-Hawthorn} \& {Gerhard}(2016)}]{2016ARA&A..54..529B}
{Bland-Hawthorn}, J. \& {Gerhard}, O. 2016, Annual Review of Astronomy and
  Astrophysics, 54, 529

\bibitem[{{Bovy} \& {Rix}(2013)}]{2013ApJ...779..115B}
{Bovy}, J. \& {Rix}, H.-W. 2013, \apj, 779, 115

\bibitem[{{Boyarsky} {et~al.}(2019){Boyarsky}, {Drewes}, {Lasserre}, {Mertens},
  \& {Ruchayskiy}}]{2019PrPNP.104....1B}
{Boyarsky}, A., {Drewes}, M., {Lasserre}, T., {Mertens}, S., \& {Ruchayskiy},
  O. 2019, Progress in Particle and Nuclear Physics, 104, 1

\bibitem[{{Boyarsky} {et~al.}(2018){Boyarsky}, {Iakubovskyi}, {Ruchayskiy}, \&
  {Savchenko}}]{2018arXiv181210488B}
{Boyarsky}, A., {Iakubovskyi}, D., {Ruchayskiy}, O., \& {Savchenko}, D. 2018,
  arXiv e-prints, arXiv:1812.10488

\bibitem[{{Boyarsky} {et~al.}(2008{\natexlab{a}}){Boyarsky}, {Iakubovskyi},
  {Ruchayskiy}, \& {Savchenko}}]{2008MNRAS.387.1361B}
{Boyarsky}, A., {Iakubovskyi}, D., {Ruchayskiy}, O., \& {Savchenko}, V.
  2008{\natexlab{a}}, \mnras, 387, 1361

\bibitem[{{Boyarsky} {et~al.}(2014){Boyarsky}, {Ruchayskiy}, {Iakubovskyi}, \&
  {Franse}}]{2014PhRvL.113y1301B}
{Boyarsky}, A., {Ruchayskiy}, O., {Iakubovskyi}, D., \& {Franse}, J. 2014,
  \prl, 113, 251301

\bibitem[{{Boyarsky} {et~al.}(2008{\natexlab{b}}){Boyarsky}, {Ruchayskiy}, \&
  {Markevitch}}]{2008ApJ...673..752B}
{Boyarsky}, A., {Ruchayskiy}, O., \& {Markevitch}, M. 2008{\natexlab{b}}, \apj,
  673, 752

\bibitem[{{Boyarsky} {et~al.}(2009){Boyarsky}, {Ruchayskiy}, \&
  {Shaposhnikov}}]{2009ARNPS..59..191B}
{Boyarsky}, A., {Ruchayskiy}, O., \& {Shaposhnikov}, M. 2009, Annual Review of
  Nuclear and Particle Science, 59, 191

\bibitem[{{Bulbul} {et~al.}(2016){Bulbul}, {Markevitch}, {Foster}, {Miller},
  {Bautz}, {Loewenstein}, {Randall}, \& {Smith}}]{2016ApJ...831...55B}
{Bulbul}, E., {Markevitch}, M., {Foster}, A., {et~al.} 2016, \apj, 831, 55

\bibitem[{{Bulbul} {et~al.}(2014){Bulbul}, {Markevitch}, {Foster}, {Smith},
  {Loewenstein}, \& {Randall}}]{2014ApJ...789...13B}
{Bulbul}, E., {Markevitch}, M., {Foster}, A., {et~al.} 2014, \apj, 789, 13

\bibitem[{{Cappelluti} {et~al.}(2018){Cappelluti}, {Bulbul}, {Foster},
  {Natarajan}, {Urry}, {Bautz}, {Civano}, {Miller}, \&
  {Smith}}]{2018ApJ...854..179C}
{Cappelluti}, N., {Bulbul}, E., {Foster}, A., {et~al.} 2018, \apj, 854, 179

\bibitem[{{Conlon} {et~al.}(2017){Conlon}, {Day}, {Jennings}, {Krippendorf}, \&
  {Rummel}}]{2017PhRvD..96l3009C}
{Conlon}, J.~P., {Day}, F., {Jennings}, N., {Krippendorf}, S., \& {Rummel}, M.
  2017, \prd, 96, 123009

\bibitem[{{Dessert} {et~al.}(2018){Dessert}, {Rodd}, \&
  {Safdi}}]{2018arXiv181206976D}
{Dessert}, C., {Rodd}, N.~L., \& {Safdi}, B.~R. 2018, arXiv e-prints,
  arXiv:1812.06976

\bibitem[{{Dodelson} \& {Widrow}(1994)}]{1994PhRvL..72...17D}
{Dodelson}, S. \& {Widrow}, L.~M. 1994, \prl, 72, 17

\bibitem[{{Foster} {et~al.}(2012){Foster}, {Ji}, {Smith}, \&
  {Brickhouse}}]{2012ApJ...756..128F}
{Foster}, A.~R., {Ji}, L., {Smith}, R.~K., \& {Brickhouse}, N.~S. 2012, \apj,
  756, 128

\bibitem[{{Franse} {et~al.}(2016){Franse}, {Bulbul}, {Foster}, {Boyarsky},
  {Markevitch}, {Bautz}, {Iakubovskyi}, {Loewenstein}, {McDonald}, {Miller},
  {Randall}, {Ruchayskiy}, \& {Smith}}]{2016ApJ...829..124F}
{Franse}, J., {Bulbul}, E., {Foster}, A., {et~al.} 2016, \apj, 829, 124

\bibitem[{{Fruscione} {et~al.}(2006){Fruscione}, {McDowell}, {Allen},
  {Brickhouse}, {Burke}, {Davis}, {Durham}, {Elvis}, {Galle}, {Harris},
  {Huenemoerder}, {Houck}, {Ishibashi}, {Karovska}, {Nicastro}, {Noble},
  {Nowak}, {Primini}, {Siemiginowska}, {Smith}, \&
  {Wise}}]{2006SPIE.6270E..1VF}
{Fruscione}, A., {McDowell}, J.~C., {Allen}, G.~E., {et~al.} 2006, in
  \procspie, Vol. 6270, Society of Photo-Optical Instrumentation Engineers
  (SPIE) Conference Series, 62701V

\bibitem[{{Garmire} {et~al.}(2003){Garmire}, {Bautz}, {Ford}, {Nousek}, \&
  {Ricker}}]{2003SPIE.4851...28G}
{Garmire}, G.~P., {Bautz}, M.~W., {Ford}, P.~G., {Nousek}, J.~A., \& {Ricker},
  George~R., J. 2003, in Society of Photo-Optical Instrumentation Engineers
  (SPIE) Conference Series, Vol. 4851, X-Ray and Gamma-Ray Telescopes and
  Instruments for Astronomy., ed. J.~E. {Truemper} \& H.~D. {Tananbaum}, 28--44

\bibitem[{{Graessle} {et~al.}(2007){Graessle}, {Evans}, {Glotfelty}, {He},
  {Evans}, {Rots}, {Fabbiano}, \& {Brissenden}}]{2007ChNew..14...33G}
{Graessle}, D.~E., {Evans}, I.~N., {Glotfelty}, K., {et~al.} 2007, Chandra
  News, 14, 33

\bibitem[{{Gu} {et~al.}(2015){Gu}, {Kaastra}, {Raassen}, {Mullen}, {Cumbee},
  {Lyons}, \& {Stancil}}]{2015A&A...584L..11G}
{Gu}, L., {Kaastra}, J., {Raassen}, A.~J.~J., {et~al.} 2015, \aap, 584, L11

\bibitem[{{Herms} {et~al.}(2018){Herms}, {Ibarra}, \&
  {Toma}}]{2018JCAP...06..036H}
{Herms}, J., {Ibarra}, A., \& {Toma}, T. 2018, Journal of Cosmology and
  Astro-Particle Physics, 2018, 036

\bibitem[{{Hitomi Collaboration} {et~al.}(2018){Hitomi Collaboration},
  {Aharonian}, {Akamatsu}, {Akimoto}, {Allen}, {Angelini}, {Audard}, {Awaki},
  {Axelsson}, {Bamba}, {Bautz}, {Blandford}, {Brenneman}, {Brown}, {Bulbul},
  {Cackett}, {Chernyakova}, {Chiao}, {Coppi}, {Costantini}, {de Plaa}, {de
  Vries}, {den Herder}, {Done}, {Dotani}, {Ebisawa}, {Eckart}, {Enoto}, {Ezoe},
  {Fabian}, {Ferrigno}, {Foster}, {Fujimoto}, {Fukazawa}, {Furuzawa},
  {Galeazzi}, {Gallo}, {Gandhi}, {Giustini}, {Goldwurm}, {Gu}, {Guainazzi},
  {Haba}, {Hagino}, {Hamaguchi}, {Harrus}, {Hatsukade}, {Hayashi}, {Hayashi},
  {Hayashida}, {Hell}, {Hiraga}, {Hornschemeier}, {Hoshino}, {Hughes},
  {Ichinohe}, {Iizuka}, {Inoue}, {Inoue}, {Ishida}, {Ishikawa}, {Ishisaki},
  {Iwai}, {Kaastra}, {Kallman}, {Kamae}, {Kataoka}, {Katsuda}, {Kawai},
  {Kelley}, {Kilbourne}, {Kitaguchi}, {Kitamoto}, {Kitayama}, {Kohmura},
  {Kokubun}, {Koyama}, {Koyama}, {Kretschmar}, {Krimm}, {Kubota}, {Kunieda},
  {Laurent}, {Lee}, {Leutenegger}, {Limousin}, {Loewenstein}, {Long}, {Lumb},
  {Madejski}, {Maeda}, {Maier}, {Makishima}, {Markevitch}, {Matsumoto},
  {Matsushita}, {McCammon}, {McNamara}, {Mehdipour}, {Miller}, {Miller},
  {Mineshige}, {Mitsuda}, {Mitsuishi}, {Miyazawa}, {Mizuno}, {Mori}, {Mori},
  {Mukai}, {Murakami}, {Mushotzky}, {Nakagawa}, {Nakajima}, {Nakamori},
  {Nakashima}, {Nakazawa}, {Nobukawa}, {Nobukawa}, {Noda}, {Odaka}, {Ohashi},
  {Ohno}, {Okajima}, {Ota}, {Ozaki}, {Paerels}, {Paltani}, {Petre}, {Pinto},
  {Porter}, {Pottschmidt}, {Reynolds}, {Safi-Harb}, {Saito}, {Sakai}, {Sasaki},
  {Sato}, {Sato}, {Sato}, {Sawada}, {Schartel}, {Serlemtsos}, {Seta},
  {Shidatsu}, {Simionescu}, {Smith}, {Soong}, {Stawarz}, {Sugawara}, {Sugita},
  {Szymkowiak}, {Tajima}, {Takahashi}, {Takahashi}, {Takeda}, {Takei},
  {Tamagawa}, {Tamura}, {Tanaka}, {Tanaka}, {Tanaka}, {Tashiro}, {Tawara},
  {Terada}, {Terashima}, {Tombesi}, {Tomida}, {Tsuboi}, {Tsujimoto}, {Tsunemi},
  {Tsuru}, {Uchida}, {Uchiyama}, {Uchiyama}, {Ueda}, {Ueda}, {Uno}, {Urry},
  {Ursino}, {Watanabe}, {Werner}, {Wilkins}, {Williams}, {Yamada}, {Yamaguchi},
  {Yamaoka}, {Yamasaki}, {Yamauchi}, {Yamauchi}, {Yaqoob}, {Yatsu}, {Yonetoku},
  {Zhuravleva}, {Zoghbi}, \& {Raassen}}]{2018PASJ...70...12H}
{Hitomi Collaboration}, {Aharonian}, F., {Akamatsu}, H., {et~al.} 2018,
  Publications of the Astronomical Society of Japan, 70, 12

\bibitem[{{Hofmann} {et~al.}(2016{\natexlab{a}}){Hofmann}, {Sanders}, {Nandra},
  {Clerc}, \& {Gaspari}}]{2016A&A...592A.112H}
{Hofmann}, F., {Sanders}, J.~S., {Nandra}, K., {Clerc}, N., \& {Gaspari}, M.
  2016{\natexlab{a}}, \aap, 592, A112

\bibitem[{{Hofmann} {et~al.}(2016{\natexlab{b}}){Hofmann}, {Sanders}, {Nandra},
  {Clerc}, \& {Gaspari}}]{2016A&A...585A.130H}
{Hofmann}, F., {Sanders}, J.~S., {Nandra}, K., {Clerc}, N., \& {Gaspari}, M.
  2016{\natexlab{b}}, \aap, 585, A130

\bibitem[{{Jeltema} \& {Profumo}(2016)}]{2016MNRAS.458.3592J}
{Jeltema}, T. \& {Profumo}, S. 2016, \mnras, 458, 3592

\bibitem[{{Kang} \& {Patra}(2016)}]{2016JKPS...69.1375K}
{Kang}, S.~K. \& {Patra}, A. 2016, Journal of Korean Physical Society, 69, 1375

\bibitem[{{Lovell} {et~al.}(2019){Lovell}, {Barnes}, {Bah{\'e}}, {Schaye},
  {Schaller}, {Theuns}, {Bose}, {Crain}, {Vecchia}, {Frenk}, {Hellwing}, {Kay},
  {Ludlow}, \& {Bower}}]{2019MNRAS.tmp..674L}
{Lovell}, M.~R., {Barnes}, D., {Bah{\'e}}, Y., {et~al.} 2019, \mnras, 674

\bibitem[{{Merle} \& {Schneider}(2015)}]{2015PhLB..749..283M}
{Merle}, A. \& {Schneider}, A. 2015, Physics Letters B, 749, 283

\bibitem[{{Merloni} {et~al.}(2012){Merloni}, {Predehl}, {Becker},
  {B{\"o}hringer}, {Boller}, {Brunner}, {Brusa}, {Dennerl}, {Freyberg},
  {Friedrich}, {Georgakakis}, {Haberl}, {Hasinger}, {Meidinger}, {Mohr},
  {Nandra}, {Rau}, {Reiprich}, {Robrade}, {Salvato}, {Santangelo}, {Sasaki},
  {Schwope}, {Wilms}, \& {German eROSITA Consortium}}]{2012arXiv1209.3114M}
{Merloni}, A., {Predehl}, P., {Becker}, W., {et~al.} 2012, arXiv e-prints,
  arXiv:1209.3114

\bibitem[{{Nakashima} {et~al.}(2013){Nakashima}, {Nobukawa}, {Uchida},
  {Tanaka}, {Tsuru}, {Koyama}, {Murakami}, \& {Uchiyama}}]{2013ApJ...773...20N}
{Nakashima}, S., {Nobukawa}, M., {Uchida}, H., {et~al.} 2013, \apj, 773, 20

\bibitem[{{Nandra} {et~al.}(2013){Nandra}, {Barret}, {Barcons}, {Fabian}, {den
  Herder}, {Piro}, {Watson}, {Adami}, {Aird}, {Afonso}, {Alexander},
  {Argiroffi}, {Amati}, {Arnaud}, {Atteia}, {Audard}, {Badenes}, {Ballet},
  {Ballo}, {Bamba}, {Bhardwaj}, {Stefano Battistelli}, {Becker}, {De Becker},
  {Behar}, {Bianchi}, {Biffi}, {B{\^\i}rzan}, {Bocchino}, {Bogdanov}, {Boirin},
  {Boller}, {Borgani}, {Borm}, {Bouch{\'e}}, {Bourdin}, {Bower}, {Braito},
  {Branchini}, {Branduardi-Raymont}, {Bregman}, {Brenneman}, {Brightman},
  {Br{\"u}ggen}, {Buchner}, {Bulbul}, {Brusa}, {Bursa}, {Caccianiga},
  {Cackett}, {Campana}, {Cappelluti}, {Cappi}, {Carrera}, {Ceballos},
  {Christensen}, {Chu}, {Churazov}, {Clerc}, {Corbel}, {Corral}, {Comastri},
  {Costantini}, {Croston}, {Dadina}, {D'Ai}, {Decourchelle}, {Della Ceca},
  {Dennerl}, {Dolag}, {Done}, {Dovciak}, {Drake}, {Eckert}, {Edge}, {Ettori},
  {Ezoe}, {Feigelson}, {Fender}, {Feruglio}, {Finoguenov}, {Fiore}, {Galeazzi},
  {Gallagher}, {Gandhi}, {Gaspari}, {Gastaldello}, {Georgakakis},
  {Georgantopoulos}, {Gilfanov}, {Gitti}, {Gladstone}, {Goosmann}, {Gosset},
  {Grosso}, {Guedel}, {Guerrero}, {Haberl}, {Hardcastle}, {Heinz}, {Alonso
  Herrero}, {Herv{\'e}}, {Holmstrom}, {Iwasawa}, {Jonker}, {Kaastra}, {Kara},
  {Karas}, {Kastner}, {King}, {Kosenko}, {Koutroumpa}, {Kraft}, {Kreykenbohm},
  {Lallement}, {Lanzuisi}, {Lee}, {Lemoine-Goumard}, {Lobban}, {Lodato},
  {Lovisari}, {Lotti}, {McCharthy}, {McNamara}, {Maggio}, {Maiolino}, {De
  Marco}, {de Martino}, {Mateos}, {Matt}, {Maughan}, {Mazzotta}, {Mendez},
  {Merloni}, {Micela}, {Miceli}, {Mignani}, {Miller}, {Miniutti}, {Molendi},
  {Montez}, {Moretti}, {Motch}, {Naz{\'e}}, {Nevalainen}, {Nicastro}, {Nulsen},
  {Ohashi}, {O'Brien}, {Osborne}, {Oskinova}, {Pacaud}, {Paerels}, {Page},
  {Papadakis}, {Pareschi}, {Petre}, {Petrucci}, {Piconcelli}, {Pillitteri},
  {Pinto}, {de Plaa}, {Pointecouteau}, {Ponman}, {Ponti}, {Porquet}, {Pounds},
  {Pratt}, {Predehl}, {Proga}, {Psaltis}, {Rafferty}, {Ramos-Ceja}, {Ranalli},
  {Rasia}, {Rau}, {Rauw}, {Rea}, {Read}, {Reeves}, {Reiprich}, {Renaud},
  {Reynolds}, {Risaliti}, {Rodriguez}, {Rodriguez Hidalgo}, {Roncarelli},
  {Rosario}, {Rossetti}, {Rozanska}, {Rovilos}, {Salvaterra}, {Salvato}, {Di
  Salvo}, {Sanders}, {Sanz-Forcada}, {Schawinski}, {Schaye}, {Schwope},
  {Sciortino}, {Severgnini}, {Shankar}, {Sijacki}, {Sim}, {Schmid}, {Smith},
  {Steiner}, {Stelzer}, {Stewart}, {Strohmayer}, {Str{\"u}der}, {Sun}, {Takei},
  {Tatischeff}, {Tiengo}, {Tombesi}, {Trinchieri}, {Tsuru}, {Ud-Doula},
  {Ursino}, {Valencic}, {Vanzella}, {Vaughan}, {Vignali}, {Vink}, {Vito},
  {Volonteri}, {Wang}, {Webb}, {Willingale}, {Wilms}, {Wise}, {Worrall},
  {Young}, {Zampieri}, {In't Zand}, {Zane}, {Zezas}, {Zhang}, \&
  {Zhuravleva}}]{2013arXiv1306.2307N}
{Nandra}, K., {Barret}, D., {Barcons}, X., {et~al.} 2013, arXiv e-prints,
  arXiv:1306.2307

\bibitem[{{Neronov} {et~al.}(2016){Neronov}, {Malyshev}, \&
  {Eckert}}]{2016PhRvD..94l3504N}
{Neronov}, A., {Malyshev}, D., \& {Eckert}, D. 2016, \prd, 94, 123504

\bibitem[{{Ng} {et~al.}(2019){Ng}, {Roach}, {Perez}, {Beacom}, {Horiuchi},
  {Krivonos}, \& {Wik}}]{2019PhRvD..99h3005N}
{Ng}, K. C.~Y., {Roach}, B.~M., {Perez}, K., {et~al.} 2019, \prd, 99, 083005

\bibitem[{{Ponti} {et~al.}(2019){Ponti}, {Hofmann}, {Churazov}, {Morris},
  {Haberl}, {Nandra}, {Terrier}, {Clavel}, \& {Goldwurm}}]{Ponti2019}
{Ponti}, G., {Hofmann}, F., {Churazov}, E., {et~al.} 2019, Nature, 567,
  347–350

\bibitem[{{Portail} {et~al.}(2017){Portail}, {Gerhard}, {Wegg}, \&
  {Ness}}]{2017MNRAS.465.1621P}
{Portail}, M., {Gerhard}, O., {Wegg}, C., \& {Ness}, M. 2017, \mnras, 465, 1621

\bibitem[{{Read}(2014)}]{2014JPhG...41f3101R}
{Read}, J.~I. 2014, Journal of Physics G Nuclear Physics, 41, 063101

\bibitem[{{Revnivtsev} {et~al.}(2009){Revnivtsev}, {Sazonov}, {Churazov},
  {Forman}, {Vikhlinin}, \& {Sunyaev}}]{2009Natur.458.1142R}
{Revnivtsev}, M., {Sazonov}, S., {Churazov}, E., {et~al.} 2009, \nat, 458, 1142

\bibitem[{{Sanders}(2006)}]{2006MNRAS.371..829S}
{Sanders}, J.~S. 2006, \mnras, 371, 829

\bibitem[{{Shah} {et~al.}(2016){Shah}, {Dobrodey}, {Bernitt},
  {Steinbr{\"u}gge}, {Crespo L{\'o}pez-Urrutia}, {Gu}, \&
  {Kaastra}}]{2016ApJ...833...52S}
{Shah}, C., {Dobrodey}, S., {Bernitt}, S., {et~al.} 2016, \apj, 833, 52

\bibitem[{{Tashiro} {et~al.}(2018){Tashiro}, {Maejima}, {Toda}, {Kelley},
  {Reichenthal}, {Lobell}, {Petre}, {Guainazzi}, {Costantini}, {Edison},
  {Fujimoto}, {Grim}, {Hayashida}, {den Herder}, {Ishisaki}, {Paltani},
  {Matsushita}, {Mori}, {Sneiderman}, {Takei}, {Terada}, {Tomida}, {Akamatsu},
  {Angelini}, {Arai}, {Awaki}, {Babyk}, {Bamba}, {Barfknecht}, {Barnstable},
  {Bialas}, {Blagojevic}, {Bonafede}, {Brambora}, {Brenneman}, {Brown},
  {Brown}, {Burns}, {Canavan}, {Carnahan}, {Chiao}, {Comber}, {Corrales}, {de
  Vries}, {Dercksen}, {Diaz-Trigo}, {Dillard}, {DiPirro}, {Done}, {Dotani},
  {Ebisawa}, {Eckart}, {Enoto}, {Ezoe}, {Ferrigno}, {Fukazawa}, {Fujita},
  {Furuzawa}, {Gallo}, {Graham}, {Gu}, {Hagino}, {Hamaguchi}, {Hatsukade},
  {Hawes}, {Hayashi}, {Hegarty}, {Hell}, {Hiraga}, {Hodges-Kluck}, {Holland},
  {Hornschemeier}, {Hoshino}, {Ichinohe}, {Iizuka}, {Ishibashi}, {Ishida},
  {Ishikawa}, {Ishimura}, {James}, {Kallman}, {Kara}, {Katsuda}, {Kenyon},
  {Kilbourne}, {Kimball}, {Kitaguti}, {Kitamoto}, {Kobayashi}, {Kohmura},
  {Koyama}, {Kubota}, {Leutenegger}, {Lockard}, {Loewenstein}, {Maeda},
  {Marbley}, {Markevitch}, {Matsumoto}, {Matsuzaki}, {McCammon}, {McNamara},
  {Miko}, {Miller}, {Miller}, {Minesugi}, {Mitsuishi}, {Mizuno}, {Mori},
  {Mukai}, {Murakami}, {Mushotzky}, {Nakajima}, {Nakamura}, {Nakashima},
  {Nakazawa}, {Natsukari}, {Nigo}, {Nishioka}, {Nobukawa}, {Nobukawa}, {Noda},
  {Odaka}, {Ogawa}, {Ohashi}, {Ohno}, {Ohta}, {Okajima}, {Okamoto}, {Onizuka},
  {Ota}, {Ozaki}, {Plucinsky}, {Porter}, {Pottschmidt}, {Sato}, {Sato},
  {Sawada}, {Seta}, {Shelton}, {Shibano}, {Shida}, {Shidatsu}, {Shirron},
  {Simionescu}, {Smith}, {Someya}, {Soong}, {Suagawara}, {Szymkowiak},
  {Takahashi}, {Tamagawa}, {Tamura}, {Tanaka}, {Terashima}, {Tsuboi},
  {Tsujimoto}, {Tsunemi}, {Tsuru}, {Uchida}, {Uchiyama}, {Ueda}, {Uno},
  {Walsh}, {Watanabe}, {Williams}, {Wolfs}, {Wright}, {Yamada}, {Yamaguchi},
  {Yamaoka}, {Yamasaki}, {Yamauchi}, {Yamauchi}, {Yanagase}, {Yaqoob},
  {Yasuda}, {Yoshioka}, {Zabala}, \& {Irina}}]{2018SPIE10699E..22T}
{Tashiro}, M., {Maejima}, H., {Toda}, K., {et~al.} 2018, in Society of
  Photo-Optical Instrumentation Engineers (SPIE) Conference Series, Vol. 10699,
  1069922

\bibitem[{{Wegg} {et~al.}(2016){Wegg}, {Gerhard}, \&
  {Portail}}]{2016MNRAS.463..557W}
{Wegg}, C., {Gerhard}, O., \& {Portail}, M. 2016, \mnras, 463, 557

\end{thebibliography}

\end{document}